\pdfoutput=1
\documentclass[aps,prl,twocolumn,showpacs,preprintnumbers,amsmath,amssymb]{revtex4-1}
\usepackage[colorlinks=true, pdfstartview=FitV, linkcolor=red, citecolor=blue, urlcolor=black, pdftitle={Counting rule for Nambu-Goldstone modes in nonrelativistic systems},pdfauthor={Yoshimasa Hidaka},pdfsubject={}, pdfkeywords={Nambu-Goldstone theorem, nonrelativistic systems, finite temperature}]{hyperref}

\usepackage{amsmath,amssymb,bm}
\usepackage[pdftex]{graphicx}
\newcommand{\tr}{\mathop{\mathrm{tr}}}
\newcommand{\tilphi}{\tilde{\phi}}
\newcommand{\tiln}{\tilde{n}}
\newcommand{\average}[1]{\langle#1\rangle}
\newcommand{\noise}{R}
\newcommand{\memory}{K}
\newcommand{\Lv}{\mathcal{L}}
\newcommand{\rank}{\mathop{\mathrm{rank}}}
\newcommand{\Potential}{\mathcal{V}}
\newcommand{\Vxx}{\Potential^{\phi\phi}}
\newcommand{\Vxp}{\Potential^{\phi n}}
\newcommand{\Vpx}{\Potential^{n\phi }}
\newcommand{\Vpp}{\Potential^{nn}}
\newcommand{\PBxx}{M_{\phi\phi}}
\newcommand{\PBxp}{{M_{\phi n}}}
\newcommand{\PBpx}{{M_{ n\phi}}}
\newcommand{\PBpp}{{M_{nn}}}
\newcommand{\two}{\text{I\hspace{-.08em}I} }
\newcommand{\NBS}{N_\text{BS}}
\newcommand{\NNG}{N_\text{NG}}
\newcommand{\NNGII}{N_\text{type-\two}}
\newcommand{\NNGI}{N_\text{type-I}}
\newcommand{\cP}{\mathcal{P}}
\newcommand{\cQ}{\mathcal{Q}}

\begin{document}

\preprint{RIKEN-MP-43}
\author{Yoshimasa Hidaka}
\affiliation{
Mathematical physics Laboratory, RIKEN Nishina Center, Saitama 351-0198, Japan}
\date{March 16, 2012}
\title{Counting rule for Nambu-Goldstone modes in nonrelativistic systems}
\begin{abstract}
The counting rule for Nambu-Goldstone modes is discussed using Mori's projection operator method in nonrelativistic systems at zero and finite temperatures.
We show that the number of Nambu-Goldstone modes is equal to the number of broken charges, $Q_a$, minus  half the rank of the expectation value of $[Q_a,Q_b]$.
\end{abstract}
\maketitle

\emph{Introduction}---
Symmetry and its spontaneous breaking are of basic importance for understanding the low-energy physics in many-body systems.
When a continuum symmetry is spontaneously broken, there exist zero modes called  Nambu-Goldstone (NG) modes~\cite{Nambu:1961tp,Goldstone:1961eq,Goldstone:1962es}, e.g.,  pions in hadron physics,  spin waves in a ferromagnet, or phonons in a crystal.
However,  the number of NG modes  associated with the spontaneous breaking appear  in a nonrelativistic system is an unsolved problem~\cite{Brauner:2010wm}. 
In the case of relativistic systems (more precisely, Lorentz invariant systems), the number of NG modes coincides with the number of broken symmetries, and their dispersion is linear, i.e., $\omega =|\bm{k}|$, where $\omega$ and $\bm{k}$ are the energy and momentum of the NG modes.
In contrast, in the case of nonrelativistic systems, the number of NG modes is not necessarily equal to the number of broken symmetries, and the dispersion may be nonlinear. 
Nielsen and Chadha (NC)~\cite{Nielsen:1975hm} classified NG modes into two types:
type-I and type-\two NG modes whose energies are odd and even powers of $|\bm{k}|$, respectively.
They showed that the number of type-I NG modes plus twice the number of type-\two NG modes is greater than or equal to the number of broken symmetries.
In known realistic examples of type-\two NG modes such as spin waves and NG modes in the Kaon condensed phase~\cite{Miransky:2001tw,Schafer:2001bq}, the inequality is saturated. 
The presence of an expectation value for charge densities implies a type-\two dispersion for NG modes, as was clarified by Leutwyler in an effective field theory~\cite{Leutwyler:1993gf}. Schafer et al. pointed out that the expectation value of $[Q_a,Q_b]$ rather than charge densities themselves plays important role,
and if they all vanish,
the number of NG modes is equal to the number of broken symmetries, where $Q_a$ are the broken charges~\cite{Schafer:2001bq}.
Recently, this theorem was generalized to the inequality,
\begin{equation}
\NBS-\NNG \leq \frac{1}{2}\rank \langle[Q_a,Q_b]\rangle,
\label{eq:WTRelation}
\end{equation}
 by Watanabe and Brauner (WB)~\cite{Watanabe:2011ec,*Watanabe:2011dk}. They conjectured that the inequality in Eq.~(\ref{eq:WTRelation}) is saturated for general systems. These results were obtained at zero temperature.

In this Letter, we generalize the Nambu-Goldstone theorem in relativistic systems to that in nonrelativistic systems at zero and finite temperatures, and
we  prove that  the inequality in Eq.~(\ref{eq:WTRelation}) is saturated.
We give an alternative definition of type-I and type-\two NG modes by using a matrix $\langle[Q_a,Q_b]\rangle$,
 and show that the number of type-\two NG modes is equal to $\rank \langle[Q_a,Q_b]\rangle/2$, and the NC inequality is also saturated.

For this purpose we apply Mori's projection operator method~\cite{Mori} to systems in which spontaneously symmetry breaking occurs. 
In this approach, a Hamilton (Langevin) equation for NG modes is derived at zero (finite) temperature, in which the expectation value of the  commutation relations for NG fields and charge densities become the Poisson brackets for their fields, i.e., NG fields and charge densities are canonical variables.
The Hamiltonian formalism for NG modes in nonrelativistic systems at zero temperature was discussed in an analysis of specific models
~\cite{Nambu:2004}.
We develop the analysis in a model independent and nonperturbative way.

In the following, we consider the case that the translation symmetry is not spontaneously broken, and
that the broken charges do not explicitly depend on spacetime variables, i.e., 
the conserved charge densities are {\it uniform}~\cite{Watanabe:2011ec},
$
n_a(t,\bm{x})=e^{iP\cdot x}n_a(0,\bm{0})e^{-iP\cdot x}
$,
where $P^\mu$ are the spacetime-translation operators.

\emph{Mori's projection operator method}---
Here, we briefly review Mori's projection operator method~\cite{Mori},
which is a powerful tool for discussing the low-energy excitations, and essential for proving our theorem for nonrelativistic systems.
For readers who are unfamiliar with this method, see, e.g., Ref.~\cite{Nordholm,Zwanzig,Rau:1995ea,Balucani2003409,Minami:2012hs} for more details of this method.

Consider a set of operators $\{A_n(t,\bm{x})\}$ corresponding to the soft modes, chosen to be NG fields and broken charge densities. 
First, we define the thermal average of an arbitrary operator $\mathcal{O}$ as
$\average{\mathcal{O}}
\equiv\tr \rho_\text{eq}\mathcal{O},
$
where we choose $\rho_\text{eq}$ as the canonical-ensemble-density operator, $\rho_\text{eq} \equiv \exp(-\beta H) / \tr \exp(-\beta H)$,
with the Hamiltonian $H$ and the inverse temperature $\beta=1/T$.
One may consider the grand canonical ensemble, by replacing $H$ with $H-\mu N$, where $\mu$ and $N$ are the chemical potential and the conserved charge operator, respectively.

Next, we introduce an inner product  for arbitrary operators $\mathcal{O}_1$ and $\mathcal{O}_2$:
\begin{equation}
(\mathcal{O}_1,\mathcal{O}_2)  \equiv \frac{1}{\beta}\int_0^\beta d\tau \average{e^{\tau H}\mathcal{O}_1e^{-\tau H} \mathcal{O}_2^\dag},
\label{eq:innerProduct}
\end{equation}
which satisfies  positive definiteness, $(\mathcal{O}_1,\mathcal{O}_1)\geq0 $, and conjugate symmetry, $(\mathcal{O}_1,\mathcal{O}_2) =(\mathcal{O}_2,\mathcal{O}_1)^* $.
Using this inner product, we define a metric:
\begin{equation}
g_{nm}(\bm{x}-\bm{y}) \equiv(A_n(0,\bm{x}), A_m(0,\bm{y})).
\end{equation}
We also define $g^{ml}(\bm{y}-\bm{z})$ as the inverse of $g_{nm}(\bm{x}-\bm{y})$, i.e.,
\begin{equation}
\int d^3yg_{nm}(\bm{x}-\bm{y})g^{ml}(\bm{y}-\bm{z})={\delta_{n}}^{l}\delta^{(3)}(\bm{x}-\bm{y}).
\end{equation}
Here, we used Einstein notation:
if an index appears twice in a single term, once as a superscript and once as subscript, 
a summation is assumed over all of its possible values.
An operator with an upper index is defined as
\begin{equation}
A^{n}(t,\bm{x}) \equiv \int d^3y g^{nm}(\bm{x}-\bm{y})A_{m}(t,\bm{y}).
\end{equation}
Since the metric is  the two-point function of $A_n$,
the inverse metric coincides with the second derivative of the effective action, $\beta\varGamma(A_n)$, with respect to $A_n$:
\begin{equation}
\begin{split}
g^{ml}(\bm{x}-\bm{y})=\frac{\delta^2 \beta\varGamma(A_n)}{\delta A_l(\bm{y}) \delta A^\dag_m(\bm{x})},
\end{split}
\end{equation}
where the effective action is given by the Legendre transformation of the generating functional $W(J^n)$:
\begin{equation}
\begin{split}
\varGamma(A_n)=  W(J^n)- \int d^3x J^m(\bm{x})\frac{\delta W(J^n)}{\delta J^m(\bm{x})} , \label{eq:effectiveAction}
\end{split}
\end{equation}
with
\begin{equation}
\begin{split}
e^{-\beta W(J^n)}=   \tr\exp \left[-\beta H+ \int d^3x A_n(0,\bm{x})J^n(\bm{x}) \right] .
\end{split}
\end{equation}

In order to decompose the degrees of freedom into the slow modes and others,
we introduce the projection operator $\cP$ acting on a field $B$, given by
\begin{equation}
\cP B(t,\bm{x}) \equiv \int d^3y  A_n(0,\bm{x}) (B(t,\bm{y}),A^n(0,\bm{x})).
\end{equation}
We also define $\cQ \equiv1-\cP$. These satisfy  $\cP^2=\cP$, $\cQ^2=\cQ$, and $\cQ\cP=\cP\cQ=0$. 

Now, let us derive a generalize Langevin equation from Mori's projection operator method by reconstructing 
 the Liouville equation,  $\partial_0 A_n(t,\bm{x})=i\Lv A_n(t,\bm{x})$, where $\Lv\mathcal{O}\equiv [H,\mathcal{O}]$.
The formal solution of the Liouville equation is obtained as $A_n(t,\bm{x})=e^{i\Lv t}A_{n}(0,\bm{x})$.

First, we decompose $\partial_0 e^{i \Lv t}$ into
\begin{equation}
\partial_0 e^{i \Lv t} = e^{i \Lv t}i\Lv= e^{i \Lv t} \cP i\Lv +e^{i \Lv t}  \cQ i\Lv. \label{eq:oi2}
\end{equation}
Next, consider the Laplace transform of $\exp({i \Lv t})$, which can be decomposed into
\begin{equation}
\frac{1}{z - i \Lv } = \frac{1}{z -\cQ i\Lv } + \frac{1}{z - i \Lv} \cP i\Lv \frac{1}{z -\cQ i\Lv }.
\end{equation}
The inverse Laplace transform leads
\begin{equation}
e^{i \Lv t}=e^{\cQ i \Lv t}+\int_{0}^{t}ds e^{i \Lv (t-s)} P i\Lv e^{\cQ i \Lv s} \label{eq:oi3}.
\end{equation}
Substituting Eq.~(\ref{eq:oi3}) into the second term in the right hand side of Eq.~(\ref{eq:oi2}), 
we obtain the operator identity,
\begin{equation}
\partial_0 e^{i \Lv t}=e^{i \Lv t} \cP i \Lv
+\int^t_0ds e^{i\Lv (t-s)} \cP i \Lv  e^{ \cQ i \Lv s} \cQ i \Lv
+e^{ \cQ i \Lv t} \cQ i \Lv. \label{eq:operatoridentity}
\end{equation}
Acting Eq.~(\ref{eq:operatoridentity}) to $A_n(0,\bm{x})$, 
we obtain the generalized Langevin equation of motion~\cite{Mori},
\begin{equation}
\begin{split}
\partial_t A_n(t,\bm{x})&=\int d^3y i{\varOmega_n}^m(\bm{x}-\bm{y})A_m(t,\bm{y})\\
&\quad-\int_0^\infty ds\int d^3y {\memory_n}^m(t-s,\bm{x}-\bm{y}) A_m(s,\bm{y})\\
&\quad+\noise_n(t,\bm{x}).
\label{eq:moriProjection}
\end{split}
\end{equation}
Here $i{\varOmega_{n}}^m(\bm{x}-\bm{y})$ is the streaming term,
${\memory_n}^m(t-s,\bm{x}-\bm{y})$ is the memory function, and $\noise_n(t,\bm{x})$ is the noise term defined as
\begin{align}
i{\varOmega_{n}}^m(\bm{x}-\bm{y})&\equiv(i\Lv A_n(0,\bm{x}),A^m(0,\bm{y}))\notag\\
&=-\frac{i}{\beta}\average{ [A_n(0,\bm{x}),A^{m\dag}(0,\bm{y})]} , \label{eq:streaming}\\
{\memory_n}^m(t-s,\bm{x}-\bm{y})&\equiv-\theta(t-s)(i\Lv \noise_n(t,\bm{x}), A^m(s,\bm{y})),\\
\noise_n(t,\bm{x})&\equiv e^{it\cQ\Lv}\cQ i\Lv A_n(0,\bm{x}), \label{eq:Rn}
\end{align}
where $\theta(t)$ is the step function.
Note that this equation is an operator identity, and thus Eq.~(\ref{eq:moriProjection}) is equivalent to the Liouville equation.
The noise term is orthogonal to $A_m(0,\bm{x})$, i.e., $(A_m(0,\bm{x}),\noise_n(t,\bm{y}))=0$,
so that it drops out when the two point function $(A_n(t,\bm{x}),A_m(0,\bm{y}))$ is considered.
Therefore, the dispersion relation for $A_n$ can be determined by the linear equation
with $i{\varOmega_n}^m(\bm{x}-\bm{y})$ and ${\memory_n}^m(t,\bm{x}-\bm{y})$.
In momentum  space, we can write the equation as
\begin{equation}
\begin{split}
\partial_t A(t,\bm{k}) &= i\varOmega(\bm{k})  A(t,\bm{k})\\
&\quad -\int ds\memory(t-s,\bm{k})A(s,\bm{k}) +R(t,\bm{k}),
\label{eq:moriProjectionInMomentumSpace}
\end{split}
\end{equation}
in matrix notation.
This is the basic equation used in the our analysis.

\emph{Nambu-Goldstone theorem}---
Let us start with the Nambu-Goldstone theorem for the effective potential~\cite{Goldstone:1962es}.
Consider a set of conserved charges,
\begin{equation}
\begin{split}
Q_a = \int d^3x \, n_a(t,\bm{x}),
\end{split}
\end{equation}
which commutes with the Hamiltonian: $[Q_a,H]=0$.
When the symmetry is spontaneously broken, there exists a matrix,
\begin{equation}
-i\average{[Q_a, \phi_i(t,\bm{x})]} \equiv [M_{n\phi}]_{ai},
\end{equation}
such that  $M_{n\phi}$ is regular,
where $i$ and $a$ run from $1$ to the number of broken symmetries $\NBS$. 
We assume that $\phi_i(x)$  are not conserved charges.
Without loss of generality, $\phi_i$ can be chosen to be a real field.

Here, let us consider the effective potential defined as $\Potential(\varPhi_i,N_a)\equiv\varGamma(A_n)/V$ by choosing $A_n=(\varPhi_i, N_a)$ that 
contain NG fields and broken charge densities, where $V$ denotes the space volume.
We parametrize $\varPhi=(\phi_i,\varphi_j)$ and $N=(n_a,n'_b)$, where   $\phi_i$ ($\varphi_j$) and $n_a$ ($n'_b$) are (non-) NG fields and (un-) broken charge densities.

The effective potential is invariant under 
the infinitesimal symmetry transformations:
\begin{align}
\delta \varPhi_i&=-i\epsilon^a \langle[Q_a, \varPhi_i]\rangle \equiv\epsilon^a [M_{N\varPhi}]_{ai}, \\
\delta N_b&=-i\epsilon^a\langle [Q_a, N_b]\rangle \equiv \epsilon^a [M_{NN}]_{ab},
\end{align}
that is, 
\begin{equation}
 [M_{N\varPhi}]_{ai}
\frac{\delta\Potential}{\delta \varPhi_i}  
+ [M_{NN}]_{ac}
\frac{\delta\Potential}{\delta N_c}  
=0.
\label{eq:identity}
\end{equation}
This is an identity of the effective potential. If $\varPhi_i$ and $N_a$ belong to linear representations of a Lie group, they become $[M_{N\varPhi}]_{ai}=-i[T_a\varPhi]_i$, and $ [M_{NN}]_{ac}= {f_{ab}}^cN_c$, where $T_a$ and ${f_{ab}}^c$ are the generator and the structure constant, respectively.
Differentiating Eq.~(\ref{eq:identity}) with respect to  $\varPhi_i$ or $N_a$, we obtain

\begin{align}
 \frac{\delta  [M_{N\varPhi}]_{ai}}{\delta \varPhi_j}
\frac{\delta\Potential}{\delta \varPhi_i}  
+  [M_{N\varPhi}]_{ai}
\frac{\delta^2\Potential}{\delta \varPhi_i\delta \varPhi_j}  \notag\\
\quad+ [M_{NN}]_{ac}
\frac{\delta^2\Potential}{\delta N_c\delta \varPhi_j} 
 &=0,  \label{eq:identity2}\\
 \frac{\delta [M_{NN}]_{ac}}{\delta N_b}\frac{\delta\Potential}{\delta N_c}  +
 [M_{N\varPhi}]_{ai}\frac{\delta^2\Potential}{\delta\varPhi_i \delta N_b}\notag\\
\quad+ [M_{NN}]_{ac}\frac{\delta^2\Potential}{\delta N_c\delta N_b}  
&=0.   \label{eq:identity3}
\end{align}
At the stationary point,  the first terms in Eqs.~(\ref{eq:identity2}) and (\ref{eq:identity3}) drop,
and only terms with NG fields and broken charges survive.
Equations~(\ref{eq:identity2}) and (\ref{eq:identity3}) become
\begin{equation}
\PBpx \Vxx+\PBpp \Vpx
=0, \quad\label{eq:NGTheorem}
\PBpx \Vxp+\PBpp\Vpp=0,
\end{equation}
where we used matrix notation, and
$\Potential^{\alpha\beta}\equiv\delta^2\Potential/(\delta\alpha\delta\beta)$ with $\alpha,\beta=(\phi_i,n_a)$.
Therefore, one finds that $(\phi^{(b)}_i,n^{(b)}_a)=([\PBxp]_{ib},[\PBpp]_{ab})$ are eigenvectors of $\Potential^{\alpha\beta}$ with the vanishing eigenvalue.
The number of eigenvectors is equal to the number of broken symmetries.
In Lorentz invariant systems, $\PBpp=0$ because non Lorentz scalar operators cannot condense. 
In this case, the inverse of propagator at $k_\mu=0$, $\Potential^{\phi\phi}$, vanishes from Eq.~(\ref{eq:NGTheorem}), which implies the propagator of $\phi$ have poles at $k^2=0$.
The number of NG modes coincides with the number of eigenvectors, i.e., the number of broken symmetries, and their dispersion is $\omega=|\bm{k}|$. This is the NG theorem in Lorentz invariant systems.
In the case of nonrelativistic systems, however,  the number of NG modes may differ from the number of broken symmetries. 

To derive the counting rule for NG modes in nonrelativistic systems, we choose the operators in Eq.~(\ref{eq:moriProjection}) as
$A_n(x)= \bigl(\tilphi_i(x),\tiln_a(x) \bigr)$ with $\tilphi_i(x)\equiv \phi_i(x)-\average{\phi_i(x)}$
and $\tiln_a(x)\equiv n_a(x) - \average{n_a(x)}$. We assume that no other zero modes that couple to NG modes exist in the system.
If this is not the case, we must add a field coupled to the zero mode to $A_n$ as a slow variable and solve the coupled equations.

Now, let us consider the Laplace transformation of the memory function, $\memory(z,\bm{k})$, which gives the dissipation of NG modes.
We are interested in the behavior of the memory function at low energy and low momentum, so we take $z\to0$ and $\bm{k}\to\bm{0}$.
Since $\mathcal{L}\tiln^a(z,\bm{k})=-\bm{k}\cdot \bm{j}^a(z,\bm{k})$, where $ \bm{j}^a(z,\bm{k})$ is the current operator,
$\memory_{nn}(z,\bm{k})\sim\bm{k}^2$, $\memory_{n\phi}(z,\bm{k})\sim\bm{k}$,
they vanish at $\bm{k}=\bm{0}$. Only $\memory_{\phi\phi}(z,\bm{k})$ can survive at $\bm{k}=\bm{0}$. 
$\memory_{\phi\phi}(z,\bm{0})$ can be expanded as
\begin{equation}
\beta \memory_{\phi\phi}(z,\bm{0})= \delta M_{\phi\phi}+L_{\phi\phi}+ z\delta Z_{\phi\phi} +\mathcal{O}(z^2).
\label{eq:Kexpand}
\end{equation}
The first and second terms $\delta M_{\phi\phi}$ and $L_{\phi\phi}$, is called the Onsager coefficients.
Both are real, and  $\delta M_{\phi\phi}^T=-\delta M_{\phi\phi}$ and  $L_{\phi\phi}^T=L_{\phi\phi}^T$ are satisfied.
$\delta M_{\phi\phi}$ gives a correction of $M_{\phi\phi}$.
$L_{\phi\phi}$, which  represents dissipation of the field  is related to a spectral function at the zero frequency, and it vanishes at zero temperature because there is no spectrum at the zero frequency
and the zero momentum. 
Since $M_{\phi\phi}(\bm{k})$ and $K_{\phi\phi}(z,\bm{k})$ appear as a combination $M_{\phi\phi}(\bm{k})-K_{\phi\phi}(z,\bm{k})$ in the Langevin equation [see Eq.~(\ref{eq:moriProjectionInMomentumSpace})],
$\delta M_{\phi\phi}$ can be renormalized to $M_{\phi\phi}$ by the shift $M_{\phi\phi}\to M_{\phi\phi}-\delta M_{\phi\phi}$ and $\beta K_{\phi\phi}\to \beta K_{\phi\phi}-\delta M_{\phi\phi}$.
The third term $\delta Z_{\phi\phi}$ denotes the correction of the wave function. 
 In the following we drop $\delta Z_{\phi\phi}$, which will be justified later in the analysis of zero modes.
First let us consider the case at zero temperature dropping the memory function. Then, the equation of motion becomes
\begin{equation}
\begin{split}
\partial_0
\begin{pmatrix}
\tilphi(t,\bm{k}) \\
\tiln(t,\bm{k})
\end{pmatrix}
&=
\begin{pmatrix}
i{\varOmega_{\phi}}^{\phi}(\bm{k})& i{\varOmega_\phi}^{ n}(\bm{k})\\
i{\varOmega_{n}}^{\phi}(\bm{k})&i{\varOmega_{n}}^n(\bm{k})
\end{pmatrix}
\begin{pmatrix}
\tilphi(t,\bm{k}) \\
\tiln(t,\bm{k})
\end{pmatrix},
\end{split}
\label{eq:HamiltonEquation}
\end{equation}
with $i{\varOmega_{\alpha}}^{\beta}(\bm{k})= M_{\alpha\gamma}(\bm{k})\varGamma^{\gamma\beta}(\bm{k})$,
where $M_{\alpha\beta}(\bm{k})$ and $\varGamma^{\alpha\beta}(\bm{k})$ is  defined as the Fourier transform of $-i\average{ [\alpha(0,\bm{x}),\beta(0,\bm{0})]}$ and $\delta^2\varGamma/(\delta\alpha(\bm{x})\delta\beta(\bm{0}))$,
which coincides with $M_{\alpha\beta}$  and $\Potential^{\alpha\beta}$ at $\bm{k}=\bm{0}$, respectively.
Equation~(\ref{eq:HamiltonEquation}) has a  form of a Hamilton equation. $M_{\alpha\beta}(\bm{k})$ can be identified as the Poisson bracket for $\alpha$ and $\beta$. 
We assume $\det M_{\alpha\beta}(\bm{k})\neq0$, which ensures $\tilphi$ and $\tiln$ are independent canonical variables.
$\varGamma^{\alpha\beta}(\bm{k})$ corresponds to the hessian of the Hamiltonian, whose eigenvalues are all positive at $\bm{k}\neq\bm{0}$,
and $\NBS$ of them becomes zero at $\bm{k}=0$, i.e., $\rank(\Potential^{\alpha\beta})=\NBS$.
The eigenmodes are obtained by $\det (i\omega+i\varOmega)=0$.
$\omega$ have always real $\pm E_n$ pairs ($E_n\geq0$, $n=1,\cdots,\NBS$) since this is the Hamiltonian system with nonnegative hessian.

At $\bm{k}=\bm{0}$, $i{\varOmega_{n}}^{\phi}(\bm{0})$ and $i{\varOmega_{n}}^{n}(\bm{0})$ vanish
from Eq.~(\ref{eq:NGTheorem}),
which is equivalent to the conservation law $dQ_a/dt=0$. 
We also obtain
\begin{equation}
i{\varOmega_{\phi}}^{n}(\bm{0})
=F^{-1},
\quad
i{\varOmega_{\phi}}^{\phi}(\bm{0})
=-F^{-1}G,
\end{equation}
where  $F^{-1}\equiv(\PBxp-\PBxx(\PBpx)^{-1} \PBpp )\Vpp$ can be identified as the inverse decay constant (matrix) for the NG modes
up to a renormalization factor, and $G=\PBpp \PBxp^{-1}$.
$F^{-1}$ is a regular matrix, which can be confirmed as follows:
Consider $\Potential' =M_{\alpha\delta}\Potential^{\delta\gamma}M_{\gamma\beta}$.
Since $M_{\alpha\beta}$ is a regular matrix, $\rank (\Potential') = \rank(\Potential^{\alpha\beta})=\NBS$ is satisfied.
$\Potential'$ is explicitly obtained as
\begin{equation}
\Potential' =
\begin{pmatrix}
F^{-1}(-G\PBxx+\PBpx) &0\\
0&0
\end{pmatrix}.
\end{equation}
Then, $\rank(\Potential')=\rank(F^{-1}(-G\PBxx+\PBpx) )=\NBS$; 
therefore, $\rank(F^{-1})=\NBS$. The rank is equal to the matrix size, so that $F^{-1}$ is a regular matrix.

The equations of motion become
\begin{align}
\partial_0 \tilphi(t,\bm{0}) &= -F^{-1}G\tilphi(t,\bm{0}) + F^{-1}\tiln(t,\bm{0}),\label{eq:eomZeromomentum}\\
\partial_0 \tiln(t,\bm{0})&= 0.
\end{align}
Then, the eigenvalue equation is $\det(i\omega+i\varOmega)=(i\omega)^{\NBS}\det(i\omega -F^{-1}G)=0$.
The number of massive modes coincides with $\rank( F^{-1}G)/2=\rank( F^{-1}M_{nn}M_{\phi n}^{-1})/2=\rank(\PBpp)/2$.
Therefore, we arrive at 
\begin{equation}
\begin{split}
\NBS - \NNG&=\frac{1}{2}\rank(\PBpp) \label{eq:WBconjecture}.
\end{split}
\end{equation}
Equation~({\ref{eq:WBconjecture}}) is nothing but the WB conjecture.

At a finite temperature, the Onsager coefficient, $L_{\phi\phi}$, contributes to the equation of motion.
This effect can be included by replacing $\PBxx$ with $\PBxx-L_{\phi\phi}$.
In this case, the above counting rule does not change as long as $\det M_{\alpha\beta}\neq0$.

Next, let us classify the NG modes. 
Differentiating Eq.~(\ref{eq:eomZeromomentum}) by $\partial_0$, we obtain
\begin{equation}
\begin{split}
\partial_0^2 \tilphi(t,\bm{0}) = -F^{-1}G\partial_0\tilphi(t,\bm{0}).
\label{eq:eom}
\end{split}
\end{equation}
We define the NG fields satisfying $G\tilphi_\text{I}=0$ as type-I NG fields, and others that are linearly independent of $\tilphi_\text{I}$ as type-\two NG fields $\tilphi_\text{\two}$.
Obviously, the number of type-I NG fields is equal to $\NBS-\rank(\PBpp)$, which coincides with the number of type-I NG bosons, $\NNGI$, because $\partial_0^2\tilphi_\text{I}=0$.
On the other hand, the number of type-\two NG modes is equal to $\rank (\PBpp)/2\equiv\NNGII$.
Since $\NNG=\NNGI+\NNGII$, we find
\begin{align}
\NBS&=\NNGI+2\NNGII ,
\label{eq:NGCounting}
\end{align}
from Eq.~(\ref{eq:WBconjecture}).
This is the saturation of the NC inequality~\cite{Nielsen:1975hm}.
Equations~({\ref{eq:WBconjecture}}) and (\ref{eq:NGCounting}) are the main results in this Letter.
Here, let us discuss whether the correction of the wave function  $\delta Z_{\phi\phi}$ in Eq.~(\ref{eq:Kexpand}) is relevant to the counting rule of the zero modes.  
$\delta Z_{\phi\phi}$ changes $\partial_0^2$ to $(1-\delta Z_{\phi\phi}G^T\Vpp G)\partial_0^2$.
For the type-I NG modes, this correction is irreverent because $G\tilphi_\text{I}=0$. 
For the type-\two NG modes, the second derivative is higher order; therefore, $\delta Z_{\phi\phi}$ is irreverent to the counting rule of NG modes.

\emph{Explicit breaking  and mass formulae}---
Here we discuss the masses of the NG modes when an explicit breaking term is added.
Suppose the symmetry is explicitly broken by  an interaction term $\delta V  = \varPhi_i h^i$.
In this case, 
the contribution from the explicit breaking term is obtained from Eq.~(\ref{eq:identity2}):
\begin{equation}
\begin{split}
[{i\varOmega_n}^\phi(\bm{0})]_{ai}= -\frac{\delta {[M_{n\varPhi}]_{aj}}}{\delta \varPhi_i}  h^j,
\end{split}
\end{equation}
where we used $\delta\Potential/\delta \varPhi_i=h^i$.
Therefore, the equation of motion becomes
\begin{equation}
\partial_0^2 \tilphi(t,\bm{0}) = -F^{-1}G \partial_0 \tilphi(t,\bm{0}) + F^{-1}i{\varOmega_{n}}^\phi(\bm{0})\tilphi(t,\bm{0}).
\label{eq:eomWithh}
\end{equation}
For general cases, it is not easy to derive the dispersion relation from Eq.~(\ref{eq:eomWithh}), so here we consider a special case: 
the symmetry breaking term does not mix the type-I and the type-\two fields, so we can decompose the explicit breaking term to
$F^{-1}{i\varOmega_n}^\phi(\bm{0})=[F^{-1}i{\varOmega_{n}}^\phi]_\text{type-I}\oplus [F^{-1}i{\varOmega_{n}}^\phi]_\text{type-\two}$.
Then, the equation of motion for $\tilphi_\text{I}$ reads 
\begin{equation}
\begin{split}
\partial_0^2 \tilphi_\text{I}(t,\bm{0}) &=[F^{-1}i{\varOmega_{n}}^\phi]_\text{type-I} \tilphi_\text{I}(t,\bm{0}),
\end{split}
\end{equation}
and thus, the squared mass matrix for type-I NG modes  becomes
$m_\text{type-I}^2= -[F^{-1}i{\varOmega_{n}}^\phi]_\text{type-I}  =\mathcal{O}(h)$,
which corresponds to  a generalized Gell-mann--Oakes--Renner relation~\cite{Son:2001ff,*Son:2002ci};
$m_\pi^2=-m_q\langle\bar{\psi}\psi\rangle \chi_I^{-1}$,
where $\langle\bar{\psi}\psi\rangle$, $ \chi_{I5}^{-1}$, and $m_q$ correspond to $M_{\phi n}$, $\Vpp$, and ${i\varOmega_n}^\phi$,
respectively. 

On the other hand, the equation of motion for $\tilphi_\text{\two}$ is 
\begin{equation}
\begin{split}
F^{-1} G\partial_0\tilphi_\text{\two}(t,\bm{0}) =i[F^{-1}i{\varOmega_{n}}^\phi]_\text{type-II} \tilphi_\text{\two}(t,\bm{0}),
\end{split}
\label{eq:type-IIequation}
\end{equation}
where we dropped the $\partial_0^2\tilphi_\text{\two}(t,\bm{0})$ term because it is higher order in $h$.
Since the $\partial_0\sim h$ in Eq.~(\ref{eq:type-IIequation}), the squared mass becomes $m_\text{type-\two}^2=\mathcal{O}(h^2)$.

The NG modes of type-I and type-\two  have different type of mass formula:
the squared mass is proportional to the $h$ and $h^2$ for type-I and type-\two NG modes, respectively.
The order counting of the mass matrices for type-I and type-\two NG modes does not change even if the mixing term between 
their modes exists in $F^{-1} i{\varOmega_{n}}^\phi$~\cite{hidaka}.
For a typical case at  finite momentum, the order of the momentum dependence  
corresponds to $h\sim\bm{k}^2$, 
then the dispersion of type-I NG modes is $\omega\sim |\bm{k}|$, and 
that of type-\two NG modes is $\omega\sim \bm{k}^2$.
In this case, the classification of type-I and type-\two in this Letter is the same as that by Nielsen-Chadha.

\emph{Summary and outlook}---
We have derived the counting rule for the NG modes, Eqs.~(\ref{eq:WBconjecture}) and (\ref{eq:NGCounting}), in nonrelativistic systems at zero and finite temperatures. 
Our method is model independent; the details of the theory are reflected by the expectation value of commutation relations for the NG fields and the broken charges, 
as well as the second derivative of the effective action.
To derive the counting rule, we employed the following assumptions:
(1) The broken charges are uniform; (2) Translation symmetry is not broken; (3) $\det M_{n\phi}\neq0$ and $\det M_{\alpha\beta}\neq0$ exist.
The assumption (1) implies that the local operator does not explicitly depends on space-time variables.
The assumption (2) is necessary that NG modes are to be eigenmodes of momentum.
The assumption (3) ensures that $\tilphi$ and $\tiln$ are independent canonical variables.
In this Letter we assumed that  $\det\langle [Q_a, \phi_i]\rangle\neq0$ and $\phi_i$ are not conserved charges. 
If a set of conserved charges $Q'_a$ exist such that $\langle [Q'_a,\phi_i] \rangle = \langle[Q'_a,Q_b]\rangle= 0$ for all $\phi_i$ and $Q_b$ ($\neq Q'_a$) but $\langle[Q'_a,Q'_b]\rangle\neq0$,
$Q'_a$ are also type-\two NG fields~\footnote {An example is the Heisenberg model, in which the spin field is conserved charge, and there is no NG field that is not a conserved charge.}. 
This can be derived using the same reasoning used in this work.
The number of these modes is equal to $\rank[Q'_a,Q'_b]$/2.
Therefore, our results in Eqs.~({\ref{eq:WBconjecture}}) and (\ref{eq:NGCounting}) remain unchanged.

Our method can also apply to the systems where spacetime symmetry is spontaneously broken; this is beyond the scope of this work.

We thank H. Abuki, T. Brauner, Y. Hama, K. Hashimoto, T. Higaki, T. Kimura, M. Murata, and K. Yazaki for useful discussions.
This work was supported by JSPS KAKENHI Grant Number No.23340067.

Note added:
After finishing this work we became aware of similar work by Watanabe and Murayama \cite{Watanabe:2012hr}.
Although their analysis is different from ours, obtained results are consistent with our work.

\bibliography{ngmode}

\providecommand{\noopsort}[1]{}\providecommand{\singleletter}[1]{#1}%
\begin{thebibliography}{22}%
\makeatletter
\providecommand \@ifxundefined [1]{%
 \@ifx{#1\undefined}
}%
\providecommand \@ifnum [1]{%
 \ifnum #1\expandafter \@firstoftwo
 \else \expandafter \@secondoftwo
 \fi
}%
\providecommand \@ifx [1]{%
 \ifx #1\expandafter \@firstoftwo
 \else \expandafter \@secondoftwo
 \fi
}%
\providecommand \natexlab [1]{#1}%
\providecommand \enquote  [1]{``#1''}%
\providecommand \bibnamefont  [1]{#1}%
\providecommand \bibfnamefont [1]{#1}%
\providecommand \citenamefont [1]{#1}%
\providecommand \href@noop [0]{\@secondoftwo}%
\providecommand \href [0]{\begingroup \@sanitize@url \@href}%
\providecommand \@href[1]{\@@startlink{#1}\@@href}%
\providecommand \@@href[1]{\endgroup#1\@@endlink}%
\providecommand \@sanitize@url [0]{\catcode `\\12\catcode `\$12\catcode
  `\&12\catcode `\#12\catcode `\^12\catcode `\_12\catcode `\%12\relax}%
\providecommand \@@startlink[1]{}%
\providecommand \@@endlink[0]{}%
\providecommand \url  [0]{\begingroup\@sanitize@url \@url }%
\providecommand \@url [1]{\endgroup\@href {#1}{\urlprefix }}%
\providecommand \urlprefix  [0]{URL }%
\providecommand \Eprint [0]{\href }%
\providecommand \doibase [0]{http://dx.doi.org/}%
\providecommand \selectlanguage [0]{\@gobble}%
\providecommand \bibinfo  [0]{\@secondoftwo}%
\providecommand \bibfield  [0]{\@secondoftwo}%
\providecommand \translation [1]{[#1]}%
\providecommand \BibitemOpen [0]{}%
\providecommand \bibitemStop [0]{}%
\providecommand \bibitemNoStop [0]{.\EOS\space}%
\providecommand \EOS [0]{\spacefactor3000\relax}%
\providecommand \BibitemShut  [1]{\csname bibitem#1\endcsname}%
\let\auto@bib@innerbib\@empty
\bibitem [{\citenamefont {Nambu}\ and\ \citenamefont
  {Jona-Lasinio}(1961)}]{Nambu:1961tp}%
  \BibitemOpen
  \bibfield  {author} {\bibinfo {author} {\bibfnamefont {Y.}~\bibnamefont
  {Nambu}}\ and\ \bibinfo {author} {\bibfnamefont {G.}~\bibnamefont
  {Jona-Lasinio}},\ }\href {\doibase 10.1103/PhysRev.122.345} {\bibfield
  {journal} {\bibinfo  {journal} {Phys. Rev.}\ }\textbf {\bibinfo {volume}
  {122}},\ \bibinfo {pages} {345} (\bibinfo {year} {1961})}\BibitemShut
  {NoStop}%
\bibitem [{\citenamefont {Goldstone}(1961)}]{Goldstone:1961eq}%
  \BibitemOpen
  \bibfield  {author} {\bibinfo {author} {\bibfnamefont {J.}~\bibnamefont
  {Goldstone}},\ }\href {\doibase 10.1007/BF02812722} {\bibfield  {journal}
  {\bibinfo  {journal} {Nuovo Cim.}\ }\textbf {\bibinfo {volume} {19}},\
  \bibinfo {pages} {154} (\bibinfo {year} {1961})}\BibitemShut {NoStop}%
\bibitem [{\citenamefont {Goldstone}\ \emph {et~al.}(1962)\citenamefont
  {Goldstone}, \citenamefont {Salam},\ and\ \citenamefont
  {Weinberg}}]{Goldstone:1962es}%
  \BibitemOpen
  \bibfield  {author} {\bibinfo {author} {\bibfnamefont {J.}~\bibnamefont
  {Goldstone}}, \bibinfo {author} {\bibfnamefont {A.}~\bibnamefont {Salam}}, \
  and\ \bibinfo {author} {\bibfnamefont {S.}~\bibnamefont {Weinberg}},\ }\href
  {\doibase 10.1103/PhysRev.127.965} {\bibfield  {journal} {\bibinfo  {journal}
  {Phys. Rev.}\ }\textbf {\bibinfo {volume} {127}},\ \bibinfo {pages} {965}
  (\bibinfo {year} {1962})}\BibitemShut {NoStop}%
\bibitem [{\citenamefont {Brauner}(2010)}]{Brauner:2010wm}%
  \BibitemOpen
  \bibfield  {author} {\bibinfo {author} {\bibfnamefont {T.}~\bibnamefont
  {Brauner}},\ }\href@noop {} {\bibfield  {journal} {\bibinfo  {journal}
  {Symmetry}\ }\textbf {\bibinfo {volume} {2}},\ \bibinfo {pages} {609}
  (\bibinfo {year} {2010})},\ \Eprint {http://arxiv.org/abs/1001.5212}
  {arXiv:1001.5212 [hep-th]} \BibitemShut {NoStop}%
\bibitem [{\citenamefont {Nielsen}\ and\ \citenamefont
  {Chadha}(1976)}]{Nielsen:1975hm}%
  \BibitemOpen
  \bibfield  {author} {\bibinfo {author} {\bibfnamefont {H.~B.}\ \bibnamefont
  {Nielsen}}\ and\ \bibinfo {author} {\bibfnamefont {S.}~\bibnamefont
  {Chadha}},\ }\href {\doibase 10.1016/0550-3213(76)90025-0} {\bibfield
  {journal} {\bibinfo  {journal} {Nucl. Phys.}\ }\textbf {\bibinfo {volume}
  {B105}},\ \bibinfo {pages} {445} (\bibinfo {year} {1976})}\BibitemShut
  {NoStop}%
\bibitem [{\citenamefont {Miransky}\ and\ \citenamefont
  {Shovkovy}(2002)}]{Miransky:2001tw}%
  \BibitemOpen
  \bibfield  {author} {\bibinfo {author} {\bibfnamefont {V.~A.}\ \bibnamefont
  {Miransky}}\ and\ \bibinfo {author} {\bibfnamefont {I.~A.}\ \bibnamefont
  {Shovkovy}},\ }\href {\doibase 10.1103/PhysRevLett.88.111601} {\bibfield
  {journal} {\bibinfo  {journal} {Phys. Rev. Lett.}\ }\textbf {\bibinfo
  {volume} {88}},\ \bibinfo {pages} {111601} (\bibinfo {year} {2002})},\
  \Eprint {http://arxiv.org/abs/hep-ph/0108178} {arXiv:hep-ph/0108178}
  \BibitemShut {NoStop}%
\bibitem [{\citenamefont {Schafer}\ \emph {et~al.}(2001)\citenamefont
  {Schafer}, \citenamefont {Son}, \citenamefont {Stephanov}, \citenamefont
  {Toublan},\ and\ \citenamefont {Verbaarschot}}]{Schafer:2001bq}%
  \BibitemOpen
  \bibfield  {author} {\bibinfo {author} {\bibfnamefont {T.}~\bibnamefont
  {Schafer}}, \bibinfo {author} {\bibfnamefont {D.~T.}\ \bibnamefont {Son}},
  \bibinfo {author} {\bibfnamefont {M.~A.}\ \bibnamefont {Stephanov}}, \bibinfo
  {author} {\bibfnamefont {D.}~\bibnamefont {Toublan}}, \ and\ \bibinfo
  {author} {\bibfnamefont {J.~J.~M.}\ \bibnamefont {Verbaarschot}},\ }\href
  {\doibase 10.1016/S0370-2693(01)01265-5} {\bibfield  {journal} {\bibinfo
  {journal} {Phys. Lett.}\ }\textbf {\bibinfo {volume} {B522}},\ \bibinfo
  {pages} {67} (\bibinfo {year} {2001})},\ \Eprint
  {http://arxiv.org/abs/hep-ph/0108210} {arXiv:hep-ph/0108210} \BibitemShut
  {NoStop}%
\bibitem [{\citenamefont {Leutwyler}(1994)}]{Leutwyler:1993gf}%
  \BibitemOpen
  \bibfield  {author} {\bibinfo {author} {\bibfnamefont {H.}~\bibnamefont
  {Leutwyler}},\ }\href {\doibase 10.1103/PhysRevD.49.3033} {\bibfield
  {journal} {\bibinfo  {journal} {Phys. Rev.}\ }\textbf {\bibinfo {volume}
  {D49}},\ \bibinfo {pages} {3033} (\bibinfo {year} {1994})},\ \Eprint
  {http://arxiv.org/abs/hep-ph/9311264} {arXiv:hep-ph/9311264 [hep-ph]}
  \BibitemShut {NoStop}%
\bibitem [{\citenamefont {Watanabe}\ and\ \citenamefont
  {Brauner}(2011)}]{Watanabe:2011ec}%
  \BibitemOpen
  \bibfield  {author} {\bibinfo {author} {\bibfnamefont {H.}~\bibnamefont
  {Watanabe}}\ and\ \bibinfo {author} {\bibfnamefont {T.}~\bibnamefont
  {Brauner}},\ }\href {\doibase 10.1103/PhysRevD.84.125013} {\bibfield
  {journal} {\bibinfo  {journal} {Phys. Rev.}\ }\textbf {\bibinfo {volume}
  {D84}},\ \bibinfo {pages} {125013} (\bibinfo {year} {2011})},\ \Eprint
  {http://arxiv.org/abs/1109.6327} {arXiv:1109.6327 [hep-ph]} \BibitemShut
  {NoStop}%
\bibitem [{\citenamefont {Watanabe}\ and\ \citenamefont
  {Brauner}(2012)}]{Watanabe:2011dk}%
  \BibitemOpen
  \bibfield  {author} {\bibinfo {author} {\bibfnamefont {H.}~\bibnamefont
  {Watanabe}}\ and\ \bibinfo {author} {\bibfnamefont {T.}~\bibnamefont
  {Brauner}},\ }\href@noop {} {\bibfield  {journal} {\bibinfo  {journal} {Phys.
  Rev.}\ }\textbf {\bibinfo {volume} {D85}},\ \bibinfo {pages} {085010}
  (\bibinfo {year} {2012})},\ \Eprint {http://arxiv.org/abs/1112.3890}
  {arXiv:1112.3890 [cond-mat.stat-mech]} \BibitemShut {NoStop}%
\bibitem [{\citenamefont {Mori}(1965)}]{Mori}%
  \BibitemOpen
  \bibfield  {author} {\bibinfo {author} {\bibfnamefont {H.}~\bibnamefont
  {Mori}},\ }\href {\doibase 10.1143/PTP.33.423} {\bibfield  {journal}
  {\bibinfo  {journal} {Prog. Theor. Phys.}\ }\textbf {\bibinfo {volume}
  {33}},\ \bibinfo {pages} {423} (\bibinfo {year} {1965})}\BibitemShut
  {NoStop}%
\bibitem [{\citenamefont {Nambu}(2004)}]{Nambu:2004}%
  \BibitemOpen
  \bibfield  {author} {\bibinfo {author} {\bibfnamefont {Y.}~\bibnamefont
  {Nambu}},\ }\href {\doibase 10.1023/B:JOSS.0000019827.74407.2d} {\bibfield
  {journal} {\bibinfo  {journal} {J. Stat. Phys.}\ }\textbf {\bibinfo {volume}
  {115}},\ \bibinfo {pages} {7} (\bibinfo {year} {2004})}\BibitemShut {NoStop}%
\bibitem [{\citenamefont {Nordholm}\ and\ \citenamefont
  {Zwanzig}(1975)}]{Nordholm}%
  \BibitemOpen
  \bibfield  {author} {\bibinfo {author} {\bibfnamefont {S.}~\bibnamefont
  {Nordholm}}\ and\ \bibinfo {author} {\bibfnamefont {R.}~\bibnamefont
  {Zwanzig}},\ }\href {\doibase 10.1007/BF01012013} {\bibfield  {journal}
  {\bibinfo  {journal} {Journal of Statistical Physics}\ }\textbf {\bibinfo
  {volume} {13}},\ \bibinfo {pages} {347} (\bibinfo {year} {1975})}\BibitemShut
  {NoStop}%
\bibitem [{\citenamefont {Zwanzig}(2001)}]{Zwanzig}%
  \BibitemOpen
  \bibfield  {author} {\bibinfo {author} {\bibfnamefont {R.}~\bibnamefont
  {Zwanzig}},\ }\href@noop {} {\emph {\bibinfo {title} {{Nonequilibrium
  Stastisical Mechanics}}}}\ (\bibinfo  {publisher} {Oxford University Press},\
  \bibinfo {year} {2001})\BibitemShut {NoStop}%
\bibitem [{\citenamefont {Rau}\ and\ \citenamefont
  {Muller}(1996)}]{Rau:1995ea}%
  \BibitemOpen
  \bibfield  {author} {\bibinfo {author} {\bibfnamefont {J.}~\bibnamefont
  {Rau}}\ and\ \bibinfo {author} {\bibfnamefont {B.}~\bibnamefont {Muller}},\
  }\href@noop {} {\bibfield  {journal} {\bibinfo  {journal} {Phys. Rept.}\
  }\textbf {\bibinfo {volume} {272}},\ \bibinfo {pages} {1} (\bibinfo {year}
  {1996})},\ \Eprint {http://arxiv.org/abs/nucl-th/9505009}
  {arXiv:nucl-th/9505009 [nucl-th]} \BibitemShut {NoStop}%
\bibitem [{\citenamefont {Balucani}\ \emph {et~al.}(2003)\citenamefont
  {Balucani}, \citenamefont {Lee},\ and\ \citenamefont
  {Tognetti}}]{Balucani2003409}%
  \BibitemOpen
  \bibfield  {author} {\bibinfo {author} {\bibfnamefont {U.}~\bibnamefont
  {Balucani}}, \bibinfo {author} {\bibfnamefont {M.~H.}\ \bibnamefont {Lee}}, \
  and\ \bibinfo {author} {\bibfnamefont {V.}~\bibnamefont {Tognetti}},\ }\href
  {\doibase 10.1016/S0370-1573(02)00430-1} {\bibfield  {journal} {\bibinfo
  {journal} {Phys. Rept.}\ }\textbf {\bibinfo {volume} {373}},\ \bibinfo
  {pages} {409 } (\bibinfo {year} {2003})}\BibitemShut {NoStop}%
\bibitem [{\citenamefont {Minami}\ and\ \citenamefont
  {Hidaka}(2012)}]{Minami:2012hs}%
  \BibitemOpen
  \bibfield  {author} {\bibinfo {author} {\bibfnamefont {Y.}~\bibnamefont
  {Minami}}\ and\ \bibinfo {author} {\bibfnamefont {Y.}~\bibnamefont
  {Hidaka}},\ }\href@noop {} {\  (\bibinfo {year} {2012})},\ \Eprint
  {http://arxiv.org/abs/1210.1313} {arXiv:1210.1313 [hep-ph]} \BibitemShut
  {NoStop}%
\bibitem [{\citenamefont {Son}\ and\ \citenamefont
  {Stephanov}(2002{\natexlab{a}})}]{Son:2001ff}%
  \BibitemOpen
  \bibfield  {author} {\bibinfo {author} {\bibfnamefont {D.}~\bibnamefont
  {Son}}\ and\ \bibinfo {author} {\bibfnamefont {M.~A.}\ \bibnamefont
  {Stephanov}},\ }\href {\doibase 10.1103/PhysRevLett.88.202302} {\bibfield
  {journal} {\bibinfo  {journal} {Phys. Rev. Lett.}\ }\textbf {\bibinfo
  {volume} {88}},\ \bibinfo {pages} {202302} (\bibinfo {year}
  {2002}{\natexlab{a}})},\ \Eprint {http://arxiv.org/abs/hep-ph/0111100}
  {arXiv:hep-ph/0111100 [hep-ph]} \BibitemShut {NoStop}%
\bibitem [{\citenamefont {Son}\ and\ \citenamefont
  {Stephanov}(2002{\natexlab{b}})}]{Son:2002ci}%
  \BibitemOpen
  \bibfield  {author} {\bibinfo {author} {\bibfnamefont {D.}~\bibnamefont
  {Son}}\ and\ \bibinfo {author} {\bibfnamefont {M.~A.}\ \bibnamefont
  {Stephanov}},\ }\href {\doibase 10.1103/PhysRevD.66.076011} {\bibfield
  {journal} {\bibinfo  {journal} {Phys. Rev.}\ }\textbf {\bibinfo {volume}
  {D66}},\ \bibinfo {pages} {076011} (\bibinfo {year} {2002}{\natexlab{b}})},\
  \Eprint {http://arxiv.org/abs/hep-ph/0204226} {arXiv:hep-ph/0204226 [hep-ph]}
  \BibitemShut {NoStop}%
\bibitem [{\citenamefont {Hidaka}()}]{hidaka}%
  \BibitemOpen
  \bibfield  {author} {\bibinfo {author} {\bibfnamefont {Y.}~\bibnamefont
  {Hidaka}},\ }\href@noop {} {\bibinfo  {journal} {in preparation}\
  }\BibitemShut {NoStop}%
\bibitem [{Note1()}]{Note1}%
  \BibitemOpen
\bibfield  {journal} {  }\bibinfo {note} {An example is the Heisenberg model,
  in which the spin field is conserved charge, and there is no NG field that is
  not a conserved charge.}\BibitemShut {Stop}%
\bibitem [{\citenamefont {Watanabe}\ and\ \citenamefont
  {Murayama}(2012)}]{Watanabe:2012hr}%
  \BibitemOpen
  \bibfield  {author} {\bibinfo {author} {\bibfnamefont {H.}~\bibnamefont
  {Watanabe}}\ and\ \bibinfo {author} {\bibfnamefont {H.}~\bibnamefont
  {Murayama}},\ }\href {\doibase 10.1103/PhysRevLett.108.251602} {\bibfield
  {journal} {\bibinfo  {journal} {Phys. Rev. Lett.}\ }\textbf {\bibinfo
  {volume} {108}},\ \bibinfo {pages} {251602} (\bibinfo {year} {2012})},\
  \Eprint {http://arxiv.org/abs/1203.0609} {arXiv:1203.0609 [hep-th]}
  \BibitemShut {NoStop}%
\end{thebibliography}%
\end{document}